# The liabilities of robots.txt

Chien-Yi Chang [a,b], Xin He [c,d,*]

[a] Assistant Professor of Law, Durham Law School, Durham University, United Kingdom
[b] Assistant Professor of Computer Science, Department of Computer Science, Durham University, United Kingdom
[c] Visiting Professor (Spring 2025), Zhongnan University of Finance and Law, Wuhan, China
[d] Mok Sau-king Professor in Law, Faculty of Law, The University of Hong Kong, Hong Kong SAR

ABSTRACT

This paper explores the legal implications of violating robots.txt, a technical standard widely used by webmasters to communicate restrictions on automated access to website content. Although historically regarded as a voluntary guideline, the rise of generative AI and large-scale web scraping has amplified the consequences of disregarding robots.txt directives. While previous legal discourse has largely focused on criminal or copyright-based remedies, we argue that civil doctrines, particularly in contract and tort law, offer a more balanced and sustainable framework for regulating web robot behavior in common law jurisdictions. Under certain conditions, robots.txt can give rise to a unilateral contract or serve as a form of notice sufficient to establish tortious liability, including trespass to chattels and negligence. Ultimately, we argue that clarifying liability for robots.txt violations is essential to addressing the growing fragmentation of the internet. By restoring balance and accountability in the digital ecosystem, our proposed framework helps preserve the internet's open and cooperative foundations. Through this lens, robots.txt can remain an equitable and effective tool for digital governance in the age of AI.

## 1. Introduction

The robots.txt file[1] is a text file that webmasters use to provide instructions about their website to web robots. These robots, sometimes also referred to as crawlers or scrapers,[2] are scripts or programs that traverse the internet to perform automated tasks. Located at the root directory of a website, the robots.txt file tells these robots which parts of the site they are allowed or disallowed from accessing and indexing. Its primary purposes are to prevent overloading of site resources due to robot visits, manage the load on the website's server, and keep specific areas of the site private. The protocol for robots.txt, commonly known as the Robots Exclusion Protocol (REP), was firstly introduced in 1994 by Martijn Koster, who[3] proposed it after seeing the need to regulate how the increasing number of web robots accessed websites. Although robots.txt is not by itself recognized as a legally binding document, it has become a widely adopted technical standard for managing server loads and balancing the interests of webmasters and web crawlers. Supported by mutual benefit considerations and the voluntary compliance of major search engine companies, robots.txt represents a powerful example of how technical standards can shape digital behavior. This dual role of robots.txt, functioning as a technical guideline and a potential regulatory mechanism, places it at the intersection of technology, law, and societal norms. Lessig's concept of "code is law"[4] aptly captures this phenomenon, suggesting that software and code can regulate digital environments in much the same way legal codes govern physical spaces.

Yet, in a time with increasingly complex digital interactions, we still lack adequate understanding, leave alone the consensus, on the liabilities of robots.txt violation. Recently, the rapid development of generative AI, such as large language models (LLMs) has brought unprecedented challenges in regulating web scraping behavior. LLMs require vast amounts of data to train effectively, leading to a surge in large-scale data collection efforts. Driven by the so-called "neural scaling law,"[5] which suggests that model performance improves significantly with increased data size, many companies, particularly major players in the AI industry, have resorted to publicly available content from the internet. While this practice is not inherently problematic, the sheer scale of data required to train LLMs has led to an unwanted increase in violations of the robots.txt instructions. At its inception, the internet was designed to embody principles of openness, collaboration, and the free exchange of information.[6] However, the aggressive data collection practices employed in the development of generative AI could undermine these ideals. The increased disregard for robots.txt directives by AI developers has created tension between webmasters and data consumers. This tension can be illustrated by our analysis of the top 200 most visited websites,[7] many of which have updated their robots.txt files to explicitly prohibit access by web robots associated with LLM training. This shows growing concern among website owners about how their data is being used, and a shift in how they attempt to assert control.

* Corresponding author. Faculty of Law, The University of Hong Kong, 10/F, Cheng Yu Tung Tower, Centennial Campus, The University of Hong Kong, Pokfulam Road, Hong Kong SAR, China.
*E-mail address:* xfhe@hku.hk (X. He).

[1] Roy T Fielding, 'Maintaining Distributed Hypertext Infostructures: Welcome to Momspider's Web' (1994) 27 Computer Networks and ISDN Systems 193.
[2] 'Web Browser' (6 December 2021) <https://web.archive.org/web/20211206205907/https://webbrowsersintroduction.com/> accessed 10 April 2024.
[3] Martijn Koster and others, 'Robots Exclusion Protocol' (Internet Engineering Task Force 2022) Request for Comments RFC 9309 <https://datatracker.ietf.org/doc/rfc9309/> accessed 10 April 2024.
[4] Lawrence Lessig, 'Code Is Law' (2000) 1 Harvard magazine.

More broadly, this development also reflects a transformation in the nature and impact of automated data access. In the early days of the internet, such access was largely unregulated, and the robots.txt protocol was widely regarded as a voluntary guideline. Violations, when they occurred, were typically seen as minor or inconsequential. However, the rise of generative AI has significantly changed this landscape. The ability of LLMs to ingest vast quantities of data has dramatically increased the need of massive amount of training data and, at the same time, amplified the potential harm to individual websites and content providers. What was once a relatively benign or overlooked practice, namely, ignoring robots.txt, can now result in severe and tangible consequences, particularly for website owners who lack the resources to maintain high-performance servers capable of handling large volumes of automated traffic. A recent example[8] involves a small, seven-person startup whose website was effectively crippled by a sudden surge of automated requests from OpenAI's web robot. The volume of access resembled a denial-of-service (DDoS) attack,[9] overwhelming the company's servers, disrupting its operations, and leading to significant financial strain. Despite attempts to block the robot via robots.txt, the protocol was ignored or bypassed, which highlights the limitations of relying on a purely voluntary system for regulating web robots.

As web scraping grows more powerful and commercially consequential, so too must the legal mechanisms for governing their use. Seeing the urgent need to clarify whether willful violations of robots.txt should give rise to liability, particularly when they result in demonstrable harm, our paper argues that common law doctrines, especially those rooted in tort and contract, offer a viable path forward for holding actors accountable, and that the law must adapt to this shifting technological environment to provide clear, consistent standards of conduct. In a post generative AI environment, leaving enforcement to individuals like in the old days is no longer sufficient. A more coherent and coordinated legal framework is needed to balance openness, innovation, and accountability.

The rest of this paper is organized as follows. Section 2 first provides definitions of key technical terms and a brief overview of the robots.txt protocol. This section then presents an empirical study that highlights the challenges prompting the need to clarify the liabilities associated with violations of robots.txt. Section 3 reviews existing literature on how robots.txt is treated. We show that this topic remains highly controversial, with little consensus. In Section 4, we present our primary legal analysis, demonstrating how violations of robots.txt can be understood through the lenses of contract law and tort law. Specifically, we provide a novel theoretical analysis of robots.txt and its relationship to website terms of use and argue that robots.txt can, under specific circumstances, function as a unilateral contract.[10] We propose that, contrary to the position[11] taken in previous literature, the terms expressed in robots.txt can be incorporated into contractual agreements, either as express or implied terms, thereby creating enforceable obligations alongside terms of use to form a binding contract. Moreover, we address situations in which civil wrongs fall outside the scope of breach of contract. Building on a critique[12] of the limited applicability of trespass to chattels in cases involving unauthorized web scraping, especially where intangible harm results from disregarding robots.txt, we argue that the tort of negligence, when applied flexibly, can fill this gap and offer meaningful legal recourse. From a policy perspective, recognizing a duty of care in such cases would not only deter harmful practices but also help mitigate the inequities and broader societal risks posed by unregulated data collection. Section 5 concludes this paper.

## 2. Background

To provide a clear foundation for the legal analysis that follows, this section introduces key technical terms (Section 2.1) and concepts (Section 2.2) relevant to the operation of robots.txt and online data collection. Furthermore, in Section 2.3, we present an empirical study that analyze how robots.txt files have changed, particularly in response to advancements in generative AI, which highlights the challenges prompting the need to clarify the liabilities associated with violations of robots.txt.

### 2.1. *Definition of terms*

For the purpose of this paper, we adopt the following working definitions to clarify key technical terms that are fundamental to the legal analysis. These definitions are intended to provide consistency and avoid ambiguity, particularly given the inconsistent usage of these terms across legal scholarship and case law:

- "Website" or "Web" refers to a collection of pages accessible via a domain[13] name, designed for users to view through a browser. It is the public-facing part of an online service.[14]
- "Server" refers to the computer system that stores and delivers website content in response to user requests.[15] It handles the data and processing behind the scenes.
- "Web Crawling" refers to the automated browsing of websites, typically conducted by search engines to index publicly accessible

[5] Jared Kaplan and others, 'Scaling Laws for Neural Language Models' [2020] arXiv preprint arXiv:2001.08361.

[6] William J Drake, Cerf G Vinton and Wolfgang Kleinwächter, 'Internet Fragmentation: An Overview' (World Economic Forum 2016).

[7] 'Top Websites Ranking - Most Visited Websites In The World' (*Similarweb*) <https://www.similarweb.com/top-websites/> accessed 22 November 2024.

[8] Julie Bort, 'How OpenAI's Bot Crushed This Seven-Person Company's Website "like a DDoS Attack"' (*TechCrunch*, 10 January 2025) <https://techcrunch.com/2025/01/10/how-openais-bot-crushed-this-seven-person-companys-web-site-like-a-ddos-attack/> accessed 11 February 2025.

[9] Jelena Mirkovic and Peter Reiher, 'A Taxonomy of DDoS Attack and DDoS Defense Mechanisms' (2004) 34 ACM SIGCOMM Computer Communication Review 39.

[10] James Barr Ames, 'Two Theories of Consideration. I. Unilateral Contracts' [1899] Harvard Law Review 515.

[11] ML Boonk and others, 'Agent Exclusion on Websites', *LEA 2005* (Wolf legal publishers 2005).

[12] MHM Schellekens, 'Are Internet Robots Adequately Regulated?' (2013) 29 Computer Law & Security Review 666.

[13] For example, www.google.com is the domain name to Google's search engine website.

[14] Niels Brügger, 'Website History and the Website as an Object of Study' (2009) 11 New Media & Society 115.

[15] Nancy J Yeager and Robert E McGrath, *Web Server Technology* (Morgan Kaufmann 1996).

content. It generally respects website architecture and protocols such as robots.txt.[16]

- "Web Scraping" refers to the automated extraction of data from websites, often targets specific content for reuse or analysis.[17] It generally does not respect protocols such as robots.txt. "Screen Scraping" refers to a type of web scraping that extracts data from the visual display of a web page, often by directly reading the rendered browser screen rather than accessing raw data.[18]
- "Web Robot", "Robot" or "Bot" refers to an automated software program that performs tasks on the internet, such as crawling (crawler) or scraping (scraper).
- "Data Mining" or "Text and Data Mining" is a broader analytical process involving the discovery of patterns or insights from large datasets.[19] In this paper, we use the term only when referring to the downstream use of data, rather than the method of its collection.

### 2.2. Basics of robots.txt

The robots.txt file is a cornerstone of web robot interaction with websites, serving as a key element of the REP. The REP provides a mechanism for webmasters to communicate crawling preferences to automated agents like search engine bots. Despite its simplicity, the protocol has evolved in both implementation and interpretation, raising significant questions about its technical and legal significance.

The REP was initially proposed to address concerns about automated bots overwhelming web servers and accessing sensitive content.[20] The robots.txt file, placed in the root directory of a website, allows webmasters to specify rules for bots, such as directories or pages that should not be crawled. The format is simple, relying on a plain-text structure to declare user-agent-specific directives like "Disallow" or "Allow." Koster's original vision for REP was to foster a cooperative relationship between webmasters and automated agents. However, the protocol lacks formal enforcement mechanisms[21] and remains a de facto standard rather than a legally binding agreement. While major search engines such as Google and Bing have integrated robots.txt interpretation into their crawling algorithms,[22] smaller or malicious bots could often disregard the protocol altogether.

A recent effort to establish an official standard is the proposal of the REP[23] by the Internet Engineering Task Force (IETF). The IETF is a global, open standards organization responsible for developing and promoting technical standards that underpin the functioning of the internet.[24] Its primary focus is on ensuring the scalability, reliability, and security of internet technologies by creating voluntary technical specifications that foster interoperability across networks and devices. The IETF is not a governmental body and holds no formal legal status. It operates under the auspices of the internet Society, a nonprofit organization that supports the development and open use of the internet. As a result, the standards developed by the IETF, published as RFCs, are voluntary. There is no legal obligation for companies, organizations, or individuals to adopt them. However, their widespread adoption is driven by industry trust and the need for interoperability. Despite its lack of formal legal authority, the IETF's work carries significant weight in the tech community. This is because its standards often become de facto requirements for ensuring compatibility and functionality on the internet.

This proposed standard, published as RFC 9309, was released in September 2022. RFC 9309 is a milestone document that represents the first official standardization of the REP, codifying decades of convention and best practices. It outlines clear specifications for the syntax and behavior of robots.txt files, addressing ambiguities that had previously led to inconsistent interpretations. Key aspects of RFC 9309 include explicit definitions of allowable syntax, handling of invalid directives, and rules for interpreting user-agent and disallow lines. Notably, RFC 9309 was published at a pivotal moment in technological history, coinciding with the advent of LLMs like OpenAI's ChatGPT, which also debuted in late 2022. This timing is significant because these models rely heavily on massive datasets scraped from the internet—a process directly impacted by the guidelines and restrictions articulated in robots.txt files. The growing prevalence of AI systems and their dependence on web-crawled data has heightened the importance of a standardized REP, as it clarifies the expectations for both webmasters and automated agents.

The syntax of the robots.txt is in many ways very similar to that used in shell scripts, which are fundamental for most, if not all, computer programmers. The directives within the file specify which web robots can access which parts (if any) of the site. When a robot visits a site, it first requests the robots.txt file. If it finds the file, it reads the directives to understand which parts of the site are off-limits and which are open. In addition to the directives, the file usually begins with non-executive headers, which are commented out with a "#" symbol. These headers or comments are merely to provide information to humans reading the file. Comments in this file might explain the context, such as issues with misbehaving robots and specific rules for different robots. However, there are no rules governing what can or cannot be included in the comments. It is entirely at the discretion of each webmaster. For example, some robots.txt file[25] set rules for robots, referencing the famous Asimov's "Three Laws of Robotics,"[26] despite knowing that the commented lines will be automatically skipped by whatever robots accessing this file. Others may include advertisements[27] for new hires. Some even lay out terms of service and prohibited uses[28] in the commented headers of their robots.txt files.

### 2.3. Changes in robots.txt files amid the rise of generative AI

To illustrate robots.txt files have changed, particularly in response to advancements in generative AI, we compare current versions of robots.txt files from the top 200 most visited domains[29] internationally based on publicly available traffic rankings from Similarweb, with their archived counterparts predating the public release of ChatGPT in November 2022.

[16] Moaiad Ahmad Khder, 'Web Scraping or Web Crawling: State of Art, Techniques, Approaches and Application.' (2021) 13 International Journal of Advances in Soft Computing & Its Applications.

[17] ibid.

[18] George H Fibbe, 'Screen-Scraping and Harmful Cybertrespass After Intel' (2003) 55 Mercer L. Rev. 1011.

[19] Ming-Syan Chen, Jiawei Han and Philip S Yu, 'Data Mining: An Overview from a Database Perspective' (2002) 8 IEEE Transactions on Knowledge and data Engineering 866.

[20] Pooja Jha and others, 'Robots Exclusion Protocol' (2014) 2 Internation journal of emerging science and engineering.

[21] ibid.

[22] 'In-Depth Guide to How Google Search Works | Google Search Central | Documentation' (*Google for Developers*) <https://developers.google.com/search/docs/fundamentals/how-search-works> accessed 2 February 2025.

[23] Koster and others (n 3).

[24] D Cooper and others, 'Internet Engineering Task Force' [2008] Fremont, CA, May.

[25] Bloomberg, 'Robots.Txt' <www.bloomberg.com/robots.txt> accessed 12 April 2024.

[26] The "Three Laws of Robotics" are a set of rules devised by science fiction writer Issac Asimov in his series of fictions, that were to be followed by robots.

[27] Bloomberg (n 25).

[28] The New York Times, 'Robots.Txt' <www.nytimes.com/robots.txt> accessed 12 April 2024.

[29] 'Top Websites Ranking - Most Visited Websites In The World' (n 7).

**Table 1**
This table shows the number of domains that have explicitly disallowed various web robots in their robots.txt files. It highlights a trend of increasing restrictions, with 71 domains blocking all robots entirely. Among specific bots, OpenAI's GPTBot is the most restricted, disallowed by 43 domains, followed closely by Common Crawl's CCBot (37 domains) and Anthropic's ClaudeBot (34 domains). The data reflects heightened concerns over web scraping activities, particularly by AI-related bots.

| Disallowed User-agent(s) | Number of Domains |
|---|---|
| * | 71 |
| GPTBot | 43 |
| CCBot | 37 |
| ClaudeBot | 34 |
| Applebot-Extended | 32 |
| anthropic-ai, FacebookBot | 31 |
| Bytespider, ChatGPT-User, PerplexityBot | 30 |
| cohere-ai | 29 |
| Facebookexternalhit | 28 |
| ImagesiftBot, Twitterbot | 27 |
| Diffbot | 26 |
| Meta-ExternalAgent, Omgilibot | 23 |
| Timpibot | 21 |
| Meta-ExternalFetcher | 20 |
| Google-Extended | 19 |
| Googlebot | 12 |
| Claude-Web | 11 |
| Amazonbot | 10 |
| Omgili | 9 |
| Omgilibot | 8 |
| Scrapy, TurnitinBot, AwarioRssBot, AwarioSmartBot | 6 |
| Bingbot, Yandex, Mediapartners-Google, bingbot, AdsBot-Google, magpie-crawler, PetalBot | 5 |
| LinkedInBot, NewsNow, news-please, Applebot, Googlebot-News, Googlebot-Video, OAI-SearchBot, FriendlyCrawler, DataForSeoBot, YouBot | 4 |
| meta-externalagent, Slurp, Google-InspectionTool, Yeti, seznambot, Discordbot, DuckDuckBot, Googlebot-Image, msnbot | 3 |
| peer39_crawler, AdsBot-Google-Mobile, Quora-Bot, peer39_crawler/1.0, Anthropic-ai, YandexBot, msnbot-media, Peer39_crawler/1.0, Omgili, Magpie-crawler, AdIdxBot, Cohere-ai, Baiduspider, SeznamBot, img2dataset, uptimerobot, teoma, TelegramBot, Neevabot, Naverbot, YaK, viberbot, Screaming Frog SEO Spider, Pinterestbot, PiplBot | 2 |
| Buzzbot, sentibot, Facebot, YJ-WSC, Y!J-WSC, IAB-Tech-Lab, BLEXBot, AlphaBot, ADmantX, Google-adstxt, HaoSouSpider, Seekr, Peer39_crawler, VelenPublicWebCrawler, GoogleOther, Googlebot-image, SemrushBot, Google-CloudVertexBot, google-extended, Acunetix Web Vulnerability Scanner, Acunetix Security Scanner, YandexVideo, Botify, YisouSpider, ia_archiver, applebot, claudebot, huggingface, claritybot, yandex.com/bot, Meltwater, Teoma, baiduspider, Momentumbot, Nutch, admantx-ussy04/3.2, CriteoBot, daumoa, panscient.com, ias_crawler, Verity, ZumBot, Perplexity-ai, scoop.it, Googlebot-Mobile, OrangeBot, SentiBot, Sogou, NewsWhip-LinkedIn-Bot, Zoombot, coccocbot-web, ByteSpider, deepcrawl, rogerbot, 360Spider, OrangeBot-Collector, MJ12bot, vebidoobot, naverbot, Mail.RU_Bot, Twitterbot/1.0, StackRambler, CriteoBot/0.1 | 1 |

The methodology for data collection leverages the robust archival capabilities of the Wayback Machine.[30] Using its Application Programming Interface (API),[31] we retrieved historical snapshots of robots.txt files for each domain, ensuring that the selected timestamps corresponded to the latest available records before November 2022. This approach guarantees a consistent and reliable baseline for comparison while accounting for the variability in archival frequency across websites. To analyze the differences, we employ a line-by-line comparison of the current and historical robots.txt files. This technique systematically identifies modifications, additions, or deletions in the directives, highlighting specific adjustments made by website administrators to their robots.txt instructions.

Our results reveal that, as shown in Table 1,[32] many webmasters, in response to perceived or actual overreach, have updated their robots.txt files to explicitly ban scraping by web robots deployed by generative AI companies. Some go so far as to prohibit all access by web robots entirely. These defensive actions reflect a growing sense of mistrust and highlight the vulnerabilities of a system that relies on voluntary compliance.

The rise of such restrictions has broader societal implications. One immediate consequence is the fragmentation of the internet. As more webmasters implement barriers to access, information that was once openly available risks being sequestered, reducing the diversity of content accessible to both humans and machines. This trend is particularly detrimental in contexts where diversity in training data is critical to developing unbiased and effective AI systems.[33] By restricting access, the internet loses its function as a shared repository of knowledge and innovation, transforming into a collection of walled gardens.

Moreover, the economic impact of large-scale scraping cannot be overlooked. Smaller websites, which often lack the infrastructure to handle the high volume of requests generated by non-compliant bots, face disproportionate burdens. Excessive server load can result in increased operational costs or service interruptions,[34] further marginalizing smaller players in the online ecosystem. This deepens the digital divide, concentrating power and influence in the hands of a few dominant players, while smaller entities are left vulnerable to exploitation. Therefore, the current situation demands a response that promotes accountability and equity. Imposing liabilities on those who violate robots.txt directives is not merely a punitive measure but a necessary step to restore the cooperative balance of the internet.

## 3. Literature review

The legal treatment of the robots.txt has generated significant academic debate, with scholars divided on its role as a technical, ethical, and legal tool. This section categorizes and compares the main strands of literature across legal doctrines and interdisciplinary perspectives, highlighting both convergence and divergence in how robots.txt is conceptualized and applied. Table 2 provides a comparative analysis of these scholarly perspectives across various themes, helping to visually summarize key points of agreement and disagreement.

Much of the existing literature examines liabilities arising from hacking protected systems or databases, or from data scraping focused on particular areas or types of content. For instance, Liu’s work[35] focuses on screen scraping in the context of the Open Banking initiative, while Xiao[36] examines data misappropriation involving trade secrets. Inspired by these works, we center our discussion on the act of violating robots.txt. As will be shown, due to the specific characteristics of robots.txt, arguments provided by the existing works cannot be directly applied and new frameworks are needed. Indeed, our paper does not confine its analysis to the nature of the content being appropriated. What we are focusing is the legal liability from the interaction between a technical protocol and the users who are often equipped by the automated generative AI. As such, we complement the content-specific discussions found in Liu, Xiao, and many others.

[30] Jamie Murphy, Noor Hazarina Hashim and Peter O'Connor, 'Take Me Back: Validating the Wayback Machine' (2007) 13 Journal of Computer-Mediated Communication 60.

[31] Joshua Bloch, 'How to Design a Good API and Why It Matters', *Companion to the 21st ACM SIGPLAN symposium on Object-oriented programming systems, languages, and applications* (2006).

[32] This table provides a detailed overview of recent trends in the evolution of robots.txt directives. Since November 2022, 143 user-agents that were previously permitted to access various domains are now explicitly banned, as indicated by updates made by domain administrators to their robots.txt files. This highlights a growing trend of increased restrictions on automated access to website content. Notably, 71 domains have taken the extreme step of blocking all robots entirely by using the wildcard syntax '*' in their robots.txt files. This directive signifies a blanket prohibition, preventing any automated user-agent from accessing the site's content. Recent updates to robots.txt files also reveal a surge in specific directives targeting AI-associated robots. 43 websites now include explicit disallow clauses for user-agents "GPTBot," a web robot developed by OpenAI to collect publicly accessible data for training large language models like GPT-4 and GPT-5.

[33] Sai S Kurapati, Antonio Yaghy and Aakriti G Shukla, 'Data Bias: Ethical Considerations for Understanding Diversity in Medical Artificial Intelligence' (2025) 5 AI and Ethics 3399.

[34] Bort (n 8).

**Table 2**
A comparative analysis of scholarly perspectives on robots.txt with respect to various themes.

| Theme | Point of Agreement | Point of Divergence |
|---|---|---|
| CFAA | Robots.txt is not consistently used as a standard of authorization. | Sellars argues it should serve as a “license” to demonstrate authorized access; O’Connor warns this risks anticompetitive outcomes; Schellekens argues CFAA and trespass fail to address scraping comprehensively. Macapinlac recommends explicitly removing scraping from CFAA jurisdiction and suggests that robots.txt be permitted as a voluntary constraint, not a basis for criminalization. |
| DMCA | Courts generally reject robots.txt as a technological protection measure. | Gallo suggests it could qualify under §512(i); Bolin dispute this position. Jasiewicz argues for potential under implied license doctrine but notes legal uncertainty. |
| Contract Law | Whether robots.txt could form an enforceable contract need further clarification. | Riley and Boonk suggest terms of use are more effective; Burrow shows UK courts increasingly enforce terms of use as backdoor protection; Jasiewicz does not emphasize on terms of use; Macapinlac supports breach of contract liability as a civil alternative to criminal penalties. |
| Tort Law | Mixed interpretations; robots.txt offers notice. | Bellia questions if it is an effective measure to convey legal limitation in trespassing. Feigin critique accessing it makes trespass theory logically inconsistent; Albritton views it as consent signaling; Schellekens finds civil trespass insufficient and advocates for self-regulatory mechanisms and governmental intervention over civil tortious liability; Sobel proposes a new category of tort specific designed for scraping; |
| EU Copyright & TDM | Some EU courts recognize opt-outs via terms of use or robots.txt. | Kokoulina and Rattzén identify uncertainty; Löbling proposes robots.txt extensions for TDM purposes. |
| Sociotechnical and Ethical Role | Seen as a voluntary, ethical guideline rather than a legal norm. | Sun evaluate ethicality via adherence; Ippolito warn against legal enforceability; Thelwall call for clearer frameworks. |
| Legal Role | Complicated and evolving. | Sellars supports recognition; Bellia and Feigin see it as insufficient; Desai notes whether ignoring it violates law is an evolving question. |

Our work also benefits from that of Kerr[37] and Christensen,[38] who primarily explore the criminal liabilities tied to unauthorized access, such as hacking. We contend, however, violating robots.txt alone does not incur criminal liability under current law. Consistent with many scholars in this area, we argue that civil liability is a more appropriate framework for addressing violations of robots.txt.

In this way, our paper builds upon existing discussions focusing on different aspects of unwanted web practices or from different perspectives. To be fair to them, many of the problems discussed in this paper do not exist at the time of their writing. As such, we align with some prior conclusions while diverging from others, situating our work as a timely addition to the existing scholarship.

### 3.1. Robots.txt under the Computer Fraud and Abuse Act (CFAA)

There is broad agreement that robots.txt has not been consistently treated as a standard for authorized access under the CFAA.[39] Sellars[40] proposes that robots.txt be interpreted as a type of “license” for automated agents, providing structured guidance on what access is permitted. By contrast, O’Connor[41] argues that relying on robots.txt under the CFAA could lead to anticompetitive outcomes by allowing website operators to control public data access unilaterally. Schellekens[42] further criticizes both CFAA and the trespass framework as poorly suited to the technological nature of scraping, noting that both doctrines fail to clearly distinguish between malicious access and legitimate automation. Macapinlac[43] calls for scraping to be explicitly removed from CFAA jurisdiction and arguing that robots.txt should remain a voluntary constraint rather than criminally enforceable.

### 3.2. Robots.txt under the Digital Millennium Copyright Act (DMCA)

Scholars generally agree that courts[44] have treated robots.txt as a “technology measure” under the DMCA §512(i).[45] Gallo[46] nonetheless contends that robots.txt may soon “outlive its usefulness as a standard technical measure.” Critics also argue that its purpose and implementation do not align with statutory goals. Bolin[47] views robots.txt as problematic and argues against recognizing robots.txt as an “effective technological protection” for copyrights, as defined in the DMCA §1201

35 Han-Wei Liu, ‘Two Decades of Laws and Practice around Screen Scraping in the Common Law World and Its Open Banking Watershed Moment’ (2020) 30 Wash. Int’l LJ 28.

36 Geoffrey Xiao, ‘Data Misappropriation: A Trade Secret Cause of Action for Data Scraping and a New Paradigm for Database Protection’ (2022) 24 Colum. Sci. & Tech. L. Rev. 125.

37 Orin S Kerr, ‘Norms of Computer Trespass’ (2016) 116 Colum. L. Rev. 1143.

38 Jennie E Christensen, ‘The Demise of the CFAA in Data Scraping Cases’ (2020) 34 Notre Dame JL Ethics & Pub. Pol’y 529.

39 Computer Fraud and Abuse Act 1986.

40 Andrew Sellars, ‘Twenty Years of Web Scraping and the Computer Fraud and Abuse Act’ (2018) 24 BUJ Sci. & Tech. L. 372.

41 Michael J O’Connor, ‘The Common Law of Cyber-Trespass’ (2019) 85 Brook. L. Rev. 421.

42 Schellekens (n 12).

43 Tess Macapinlac, ‘The Legality of Web Scraping: A Proposal’ (2018) 71 Fed. Comm. LJ 399.

44 *Healthcare Advocates, Inc v Harding, Earley, Follmer & Frailey* (2007) 497 F Supp 2d 627 (ED Pa).

45 Digital Millennium Copyright Act § 512(i).

46 Lauren G Gallo, ‘The (Im) Possibility of Standard Technical Measures for UGC Websites’ (2010) 34 Colum. JL & Arts 283.

47 Rebecca Bolin, ‘Locking Down the Library: How Copyright, Contract, and Cybertrespass Block Internet Archiving’ (2006) 29 Hastings Comm. & Ent. LJ 1.

(a)[48] because it would make circumventing a robots.txt file a crime. Jasiewicz[49] adds that while the doctrine of implied license may offer a potential legal basis for interpreting robots.txt, its status remains legally uncertain.

### 3.3. Robots.txt under the EU copyright directive

In the EU, robots.txt has become increasingly relevant under Article 4(3) of the EU Copyright Directive,[50] which allows rights holders to opt out of text and data mining (TDM). Some courts[51] recognize opt-outs expressed in the terms of use or through robots.txt, but there remains legal uncertainty. Kokoulina[52] and Rattzén[53] note that courts have not settled whether machine-readable signals like robots.txt are sufficient, or whether human-readable terms are required. To address this gap, Löbling[54] proposes extending the robots.txt format to include purpose-specific exclusions (e.g., disallowing TDM but permitting indexing), creating more nuanced and enforceable opt-out protocols.

### 3.4. Robots.txt under contract law

The enforceability of robots.txt under contract law remains contested. Riley[55] and Boonk[56] argue that terms of use, especially when presented with adequate notice, provide a firmer legal basis for enforcing scraping restrictions. Burrow[57] adds that courts in the UK have increasingly upheld terms of use as enforceable "backdoor" mechanisms for restricting scraping. In contrast, Jasiewicz[58] places less emphasis on the role of terms of use and focuses on the instructions of the robots.txt itself but admit that whether these instructions alone can constitute enforceable contracts would need further clarification. Macapinlac[59] proposes that breach of contract should serve as the principal legal remedy for unauthorized scraping and violation of robots.txt, offering civil liability as a fairer and more flexible alternative to criminal enforcement.

### 3.5. Robots.txt under tort law

Tort law offers a diverse set of theories for evaluating the legal weight of robots.txt. While many scholars agree that the file serves a notice function, they diverge on whether that notice is sufficient to ground tortious liability. Bellia[60] questions whether robots.txt provides the level of clarity needed to support claims such as trespass to chattels. Feigin[61] critiques the logical basis of such claims, pointing out that a server must already have accepted a connection before a robot can read the robots.txt file, thus undermining arguments that the access was unwanted. Albritton,[62] however, sees robots.txt as a valid form of consent signaling and suggests that standardization could enhance its legal effect. Schellekens[63] argues that civil tort doctrines are generally ill-suited to address automated scraping and calls for regulatory and institutional solutions instead. Sobel[64] introduces a new angle by proposing a tort for "bad faith scraping," focused on intentional violations of platforms' rules that cause harm to users or the broader public.

### 3.6. Robots.txt's Sociotechnical and ethical role

Technical scholars often frame robots.txt not as a legal device but as a norm rooted in ethics and community practice. Sun[65] evaluate crawler behavior using ethical principles, measuring respect for robots.txt as a proxy for ethical data access. Ippolito[66] warn against legally formalizing a convention that was never meant to carry legal weight, highlighting the risk of overextending informal norms. Thelwall[67] acknowledge the ethical and legal ambiguity of robots.txt and advocate for regulatory clarification if it is to serve a binding role.

### 3.7. Robots.txt's legal role

Even among legal scholars, there is no consensus regarding robots.txt's general legal role. Desai[68] notes whether ignoring it violates law is an evolving question. Sellars[69] argues for its proper recognition under statutory and common law frameworks. Bellia[70] and Feigin,[71] however, find its notice value insufficient to trigger legal obligations.

### 3.8. Discussion

While robots.txt originated as a technical standard to guide ethical behavior among web robots, it now sits at the intersection of multiple legal domains. Criminal law, particularly under statutes like the CFAA in the US or the Computer Misuse Act[72] in the UK, remains a controversial foundation for regulating scraping, often leading to overly broad interpretations that risk anticompetitive outcomes and undermine the openness of the internet. Copyright-based claims, such as those under the DMCA or the EU Copyright Directive, similarly struggle to address

48 Digital Millennium Copyright Act (n 45) §1201(a).
49 Monika Isia Jasiewicz, 'Copyright Protection in an Opt-out World: Implied License Doctrine and News Aggregators' (2012) 122 Yale LJ 837.
50 Directive (EU) 2019/790 of the European Parliament and of the Council of 17 April 2019 on copyright and related rights in the Digital Single Market and amending Directives 96/9/EC and 2001/29/EC.
51 Jasiewicz (n 49).
52 Olga Kokoulina, 'Towards Future-Proof, Rights-Respecting Automated Data Collection: An Examination of European Jurisprudence' (2023) 26 Vand. J. Ent. & Tech. L. 707.
53 Mattias Rattzen, 'Location Is All You Need: Copyright Extraterritoriality and Where to Train Your AI' (2024) 26 Colum. Sci. & Tech. L. Rev. 175.
54 Lisa Lobling and others, 'Navigating the Legal Landscape: Technical Implementation of Copyright Reservations for Text and Data Mining in the Era of AI Language Models' (2023) 14 J. Intell. Prop. Info. Tech. & Elec. Com. L. 499.
55 Kathleen C Riley, 'Data Scraping as a Cause of Action: Limiting Use of the CFAA and Trespass in Online Copying Cases' (2018) 29 Fordham Intell. Prop. Media & Ent. LJ 245.
56 Boonk and others (n 11).
57 S Burrow, 'The Law of Data Scraping: A Review of UK Law on Text and Data Mining' [2021] REATe Working Paper 2021/2, Zenodo.
58 Jasiewicz (n 49).
59 Macapinlac (n 43).
60 Patricia L Bellia, 'Defending Cyberproperty' (2004) 79 NYUL Rev. 2164.
61 Eric J Feigin, 'Architecture of Consent: Internet Protocols and Their Legal Implications' (2003) 56 Stan. L. Rev. 901.
62 Melvin Albritton, 'Swatting Spiders: An Analysis of Spider Activity on the Internet' (2001) 3 Tul. J. Tech. & Intell. Prop. 137.
63 Schellekens (n 12).
64 Benjamin LW Sobel, 'A New Common Law of Web Scraping' (2021) 25 Lewis & Clark L. Rev. 147.
65 Yang Sun, Isaac G Councill and C Lee Giles, 'The Ethicality of Web Crawlers', *2010 IEEE/WIC/ACM International Conference on Web Intelligence and Intelligent Agent Technology* (IEEE 2010).
66 Daphne Ippolito and Yun William Yu, 'Donottrain: A Metadata Standard for Indicating Consent for Machine Learning', *Proceedings of the 40th International Conference on Machine Learning* (2023).
67 Mike Thelwall and David Stuart, 'Web Crawling Ethics Revisited: Cost, Privacy, and Denial of Service' (2006) 57 Journal of the American Society for Information Science and Technology 1771.
68 Deven R Desai and Mark Riedl, 'Between Copyright and Computer Science: The Law and Ethics of Generative aI' (2024) 22 Nw. J. Tech. & Intell. Prop. 55.
69 Sellars (n 40).
70 Bellia (n 60).
71 Feigin (n 61).
72 Computer Misuse Act 1990.

the act of unauthorized access itself, focusing instead on the nature of the content being copied.

In contrast, civil remedies rooted in contract and tort law, especially as developed in common law jurisdictions like the US, UK, and Canada, offer more proportionate, flexible, and context-sensitive mechanisms for addressing robots.txt violations. These jurisdictions share a doctrinal reliance on precedent, making it possible to meaningfully compare and adapt legal reasoning across borders. Following this line, we argue that, under certain conditions, robots.txt can form the basis of a unilateral contract, or alternatively serve as constructive notice sufficient to ground tortious liability, including trespass to chattels and negligence.

## 4. Liabilities of robots.txt

*hiQ Labs, Inc. v LinkedIn Corp.*,[73] a very recent case addressing the liabilities of web scraping, involves hiQ disregarding LinkedIn's explicit prohibition of access by bypassing its robots.txt file. Among other measures, hiQ sought an injunction from the court to prevent LinkedIn from blocking its access. This prolonged litigation has oscillated between the District Court, the Ninth Circuit, and the Supreme Court since 2019. In June 2021, based on its decision in *Van Buren v United States*,[74] the US Supreme Court vacated the District Court's initial issuance of an injunction and remanded the case to the Ninth Circuit for reconsideration. However, the Ninth Circuit reaffirmed its previous decision, ruling that the CFAA[75] does not apply to hiQ's actions. In its second opinion, the Ninth Circuit clarified that a general violation under the CFAA does not extend to entities that perceive themselves as victims of data scraping.

While the court was reluctant to grant statutory remedies under the CFAA, it noted that those who perceive themselves as victims of data scraping are "not without resorts."[76] Alternative legal avenues are available. Among others, the court suggested potential claims could be pursued for breach of contract and tortious actions. However, the court's opinion did not delve into the details of how these courses of action might be enforced.

In this section, we follow the roadmap outlined in the Ninth Circuit's second opinion and provide our analysis.

### 4.1. Contract

In Jasiewicz, the author admits that "it is unclear whether the instructions of a robots exclusion protocol can be considered a legally enforceable contract."[77] While we agree that this issue requires litigation for certainty, applying principles of contract law suggests it is more likely than not that, from the moment one party deploys a web robot to access the robots.txt file and retrieves the other party's website content, a unilateral contract is formed. Furthermore, we argue that, contrary to Boonk's claim[78] that the terms of the robots.txt cannot be considered part of a contract unless explicitly included in the terms of use, these terms can be incorporated into such a contract either as express terms or implied terms, even without direct reference in the terms of use. The following is a detailed, step-by-step analysis of the key elements of a contract: offer, acceptance, consideration, intention to be legally bound, and certainty of terms.

An offer to make a contract shall be construed as inviting acceptance in any manner and by any medium reasonable in the circumstances.[79] In the circumstances of web scraping, when the offeror (i.e. the webmaster) created and uploaded the robots.txt file onto the root directory of the website, she invites potential offeree to accept this offer by deploying web robots to access her website's content. This offer is clearly unilateral. When it comes to unilateral offer, it is the action, not an express statement of agreement, that signifies offeree's acceptance. In *Register.com, Inc. v Verio, Inc.,* the court[80] finds that an express statement of agreement was not always required in either paper or online contracts as follows:

We recognize that contract offers on the internet often require the offeree to click on an "I agree" icon. And no doubt, in many circumstances, such a statement of agreement by the offeree is essential to the formation of a contract. But not in all circumstances. While new commerce on the internet has exposed courts to many new situations, it has not fundamentally changed the principles of contract. It is standard contract doctrine that when a benefit is offered subject to stated conditions, and the offeree makes a decision to take the benefit with knowledge of the terms of the offer, the taking constitutes an acceptance of the terms, which accordingly become binding on the offeree.

When an offeree silently deploys a web robot to access the offeror's robots.txt and then the website content, an acceptance by action is given. In *Century 21 v Rogers Communications Inc.,* the court held[81] that the automated nature of web robots does not alter the legal principles governing acceptance. If the initial access was not automated and an individual programmed the automated indexing, liability cannot be avoided merely by automating the actions.

This reasoning aligns with Lord Denning's judgment in *Thornton v Shoe Lane Parking*,[82] where he held that the terms of a contract must be made available before acceptance and that acceptance occurs when the customer commits the act that triggers performance. Lord Denning famously wrote: "The offer is contained in the notice at the entrance. The acceptance takes place when the customer puts his money into the slot." The machine did not need to verify intent; the physical act of payment was sufficient to create a binding contract on the terms offered. Analogously, when a party deploys a web robot, whether it is a search engine crawler or a scraper like OpenAI's GPTBot, that accesses a site starting with robots.txt, the webmaster's instructions function as a pre-access notice, much like the posted terms in *Thornton v Shoe Lane Parking*. The bot's continued operation signals constructive acceptance of those terms through action. This is not a passive or accidental interaction: the deployment of the robot, including its programmed behavior to read and (potentially) disregard the robots.txt file, reflects a deliberate engagement with the website's access conditions.

Moreover, *Thornton v Shoe Lane Parking* reinforces that where standard-form terms are made publicly available in advance of the transaction, such as through robots.txt, an actor is bound by those terms if they proceed. Just as a customer cannot later claim ignorance of parking terms after using the machine, a bot deployer cannot claim ignorance or deny consent after initiating automated access to a site where restrictions have been clearly posted. Thus, *Thornton v Shoe Lane Parking* strengthens the analogy between contract formation in physical and digital environments. It supports the view that robots.txt, while technical in form, can operate as a mechanism for offering terms, and that initiating access through a web robot constitutes acceptance by conduct.

An interesting side note arises when considering the absence of a robots.txt file: can a web robot deployer argue that this omission constitutes an implied invitation to access the website, thereby creating a "meeting of the minds"? In other words, does the lack of an explicit prohibition suggest that the webmaster consents to automated access?

[73] *hiQ Labs, Inc v LinkedIn Corp* (2019) 938 F3d 985 (9th Cir).
[74] *Van Buren v United States* (2021) 593 US 374 (Supreme Court).
[75] Computer Fraud and Abuse Act (n 39).
[76] *hiQ Labs, Inc v LinkedIn Corp* (9th Cir) [41].
[77] Jasiewicz (n 49).
[78] Boonk and others (n 11).
[79] District of Columbia Code § 28:2–206, 'Offer and acceptance in formation of contract.'
[80] *Register.com, Inc v Verio, Inc* (2004) 356 F3d 393 (2d Cir).
[81] *Century 21 Canada Limited Partnership v Rogers Communications Inc* 2011 BCSC 1196.
[82] *Thornton v Shoe Lane Parking* (1971) 1 ER 686 (CA).

This argument was addressed and rejected in *Associated Press v Meltwater U.S. Holdings, Inc.,* where the court held that[83] "there is no fair inference, based simply on the absence of the robots.txt protocol, that there has been a meeting of the minds." The court emphasized that silence does not equal consent or acceptance. This reinforces the idea that acceptance must be affirmatively expressed, not presumed from inaction. Just as entering unfenced land in the real world does not necessarily indicate permission to trespass, the absence of robots.txt should not be construed as a blanket authorization for automated data harvesting. The burden remains on the deployer to ensure that their actions are consistent with the terms and expectations of the host website, not the other way around.

Once the offer is accepted by the offeree, there must be consideration for a contract to form. Under common law principles, consideration is defined as something of value to the promisor or a detriment to the promise.[84] *Prima facie*, there appears to be no consideration in this case since webmasters typically do not request payment in exchange for access to their website content. However, it is important to recognize that something of value to the promisor does not always need to be monetary or tangible; it can be as nominal as a peppercorn.[85] The robots' access to a website increases traffic, thereby enhancing the website's visibility and potentially generating profits for the webmaster. This is particularly evident when robots deployed by search engines index website content. By making the website accessible through the search engine, the webmaster essentially receives advertising services in exchange for allowing robots to access the content. Even if the increased traffic does not directly result in profit, as established in *Century 21 v Rogers Communications Inc.*, courts have recognized that the content created by website developers holds intrinsic value.[86] The information accessed by robots has value in and of itself. Moreover, the actions taken by the deployer of the robots—accessing and downloading content—confirm that the information has value. If the information were without value, the deployer would not expend resources to deploy robots to crawl the website in the first place. As the judge concluded: "In my opinion, there is consideration for the contract as [one] obtained the benefit of the information displayed on the website." In cases where robots crawl the internet to download content for training LLMs, it is a matter of fact that the companies deploying these robots are profiting from the information displayed on websites. For instance, OpenAI charges significant fees for access to and use of its LLMs, further underscoring the value of the information being utilized.

Even when an offer is accepted by an offeree with consideration, a contract will not be enforceable if there is no intention to create legal relations. This issue commonly arises in cases involving family members, where the parties typically lack the intention to hold each other legally accountable.[87] In contrast, when a webmaster purposefully drafts the content of a robots.txt file and uploads it to the root of their web server, the intention to be legally bound becomes apparent. This deliberate effort reflects the webmaster's aim to increase their website's visibility and, ultimately, profitability. In agreements of a commercial nature, such as this, there is a presumption that the parties intended to create legal relations. The burden of rebutting this presumption in commercial agreements is particularly heavy and rests on the party challenging the existence of such intent.[88]

Boonk argues that for an exclusion to be effective, it should be included in the terms of use rather than in the robots.txt, as the exclusion in the robots.txt cannot be considered part of the contract. We argue the opposite: the terms in the robots.txt can be incorporated into a contract, either as unsigned express terms or as implied terms. The key to incorporating unsigned express terms into a contract lies in their timeliness, the nature of the contractual document, and reasonable notice.[89] In *Thornton v Shoe Lane Parking Ltd*, it was established that[90] notice of the terms must be provided before or at the time of contracting. In this case, the terms outlined in the robots.txt are created and uploaded to the server before the web robot accesses the robots.txt. While a defendant might argue that the plaintiff did not take reasonable steps to draw the defendant's attention to the terms in the document (as in *Parker v South Eastern Railway Co*), the fact that the defendant voluntarily deployed a robot to access the robots.txt demonstrates that the defendant was already aware of its existence. It is reasonable to expect the defendant to have understood that the robots.txt contains terms related to allowing or disallowing access. The most contentious issue remains whether the terms are presented in a manner that a reasonable person would understand to have contractual effect.[91] From our analysis of contract formation, a reasonable person would understand the robots.txt to have contractual effect. However, even if future rulings determine otherwise, the terms in the robots.txt can still be incorporated into a contract as implied terms. The use of robots.txt is widely accepted within the internet community and is considered a "community norm."[92] As established in *Smith v Wilson*, established usages and local customs may be implied into a contract based on the presumption that the parties did not intend to express the entirety of their agreement in writing but rather intended to refer to known usages.[93] Therefore, the terms within the robots.txt may be incorporated into contracts even without direct reference.[94]

It is noteworthy that the above analysis is made independently of the existence of a website's terms of use. In essence, as long as the deployer of the robot accesses the robots.txt file on the webmaster's website and subsequently interacts with the rest of the site, an implied-in-fact unilateral contract arises, independent of whether terms of use exist. There should be no debate over whether terms of use can constitute a contract, as this has been firmly established by prior cases. Our analysis further clarifies that even in the absence of terms of use, the interactions between the deployer and the webmaster—specifically regarding the robots.txt file—can form a contract. In practice, webmasters are encouraged to, and often do provide both a robots.txt file and terms of use. Therefore, it is more practical to consider the robots.txt file and the terms of use as complementary components. A well-drafted terms of use document undeniably strengthens contractual claims, and we strongly recommend that all webmasters include one. When a terms of use document is present, the terms in the robots.txt file can easily be incorporated into the contract through direct reference.[95] This is also affirmed by the recent case *hiQ Labs, Inc. v LinkedIn Corp.*, where a breach of contract was established. The court found, among others, that "entities seeking to crawl LinkedIn … must abide by the terms of LinkedIn's robots.txt and LinkedIn Crawling Terms and Conditions."[96] Additionally, in the absence of such a reference, the situation reverts to the above discussion of "established usage," and the terms in the robots.txt can still be incorporated into the contract without explicit reference.

There are many reasons why courts might be reluctant to recognize robots.txt as a contract. One of the most common arguments involves

83 *Associated Press v Meltwater US Holdings, Inc* (2013) 931 F Supp 2d 537 (SDNY).

84 *Currie v Misa* (1875) 10 LR 153.

85 *Chappell & Co Ltd v The Nestlé Co Ltd* [1960] AC 87 (HL).

86 *Century 21 Canada Limited Partnership v Rogers Communications Inc.* (n 81) 123.

87 *Balfour v Balfour* (1919) 2 KB 571 (CA).

88 *Edwards v Skyways Ltd* (1964) 1 ER 494 (CA).

89 *Parker v South Eastern Railway Co* (1876) 2 LR 416 (CA).

90 *Thornton v Shoe Lane Parking* (n 82).

91 *Chapelton v Barry Urban District Council* (1940) 1 KB 532.

92 Jasiewicz (n 49) 844.

93 *Smith v Wilson* (1832) 110 ER 266 (KB).

94 *Liverpool City Council v Irwin* [1977] AC 239 (HL).

95 *Thompson v London, Midland and Scottish Railway Company* (1903) 1 KB 41 (KB).

96 *hiQ Labs, Inc v LinkedIn Corporation* [2021] ND Cal 17-cv-03301-EMC.

questioning whether the file possesses "contractual character."[97] For example, terms printed on the back of a receipt, even if the receipt is signed, cannot be considered a contract nor can those terms be incorporated into an existing contract. We argue that rejecting robots.txt on the grounds of "lacking contractual character" risks falling into a circular logic loop. A document can only have contractual character if, to a reasonable person, it is not something inherently devoid of such character. In the case of a receipt, no reasonable person would view it as a contract or expect contractual terms to be located on its reverse side. However, as discussed above, the nature of robots.txt is, at best, ambiguous and controversial.

It is also important to note that, under common law principles, contract formation does not generally require formality. When assessing potential contractual liabilities related to violating robots.txt, the key issue is not whether the robots.txt file itself constitutes a contract, but rather an evaluation of the entire context. Courts should focus on substance over form, considering the actions of both the deployer of the robots and the webmaster, the directives within the robots.txt file, the terms of use, and established industry norms and local customs. This holistic approach determines whether contractual liability might be imposed.

### 4.2. Tort

In cases where a webmaster disallows any access by web robots to any part of its website through the robots.txt file, there is no offer being made, and thus, no contract is formed. At any time, there is simply no "meeting of minds": the webmaster does not allow access by web robots, but the deployer of the robot seeks access regardless. In this situation, the robots.txt file functions as a mere notice, akin to a "no trespassing" sign. If the deployer nevertheless proceeds to deploy their robots to scrape the webmaster's website content against the webmaster's explicit intentions, what remedies are available?

As discussed in Section 2.2, Section 2.3, and in many prior works, when website content is protected by copyright law, legal action may be pursued for unauthorized use that constitutes copyright infringement. However, not all content is protected by copyright,[98] and even when it is, establishing infringement can be challenging.[99] This paper focuses instead on the liability arising from violations of robots.txt, independent of whether the content itself is protected by copyright. We therefore address a more general scenario: where the scraped content is not copyright-protected, and the civil wrong does not fall within the scope of a breach of contract, what legal recourse remains for the plaintiff?

The case of *eBay, Inc. v Bidder's Edge*[100] was a seminal case addressing such a scenario. In this case, Bidder's Edge (BE) used web robots to collect auction data from eBay, despite eBay's use of "robot exclusion headers" (robots.txt) to explicitly state that unauthorized robotic activity was not permitted. eBay argued that this constituted trespass to chattels, asserting that the bots placed undue strain on its servers and disrupted operations.[101] The court did not address whether the auction data was copyright-protected or whether there had been infringement, as these issues[102] were irrelevant to the claim of trespass to chattels. The court specifically noted: "BE is factually incorrect to the extent it argues that the trespass claim arises out of what BE does with the information it gathers by accessing eBay's computer system, rather than the mere fact that BE accesses and uses that system without authorization." Unsurprisingly, the court ruled in favor of eBay, granting a preliminary injunction against BE and recognizing that the bots' activities placed an undue load on eBay's servers. The decision emphasized that BE's bots interfered with eBay's property rights by exploiting server resources without consent, even in the absence of direct damage. While trespass to chattels traditionally involves unlawful interference with another person's tangible personal property,[103] this case set a precedent for applying the doctrine in the context of robotic access to online systems.

To establish a claim of trespass to chattels in situations where a web robot accesses a website while ignoring the rules set forth in a robots.txt file, one must consider the key elements required for such a tort claim. These elements typically include the lack of consent, interference, intention, and actual harm.[104]

The first element to establish is that the webmaster (plaintiff) did not consent to the web robot of deployer (defendant) accessing certain parts of the website. The robots.txt file plays a crucial role in this regard, serving as clear evidence of non-consent. This file explicitly specifies which parts of the website are off-limits to robots, allowing the webmaster to communicate their preferences regarding access. By identifying directories or pages that should not be accessed, the robots.txt file clearly signals a lack of consent, thereby satisfying the first criterion for the tort. It is generally presumed that operators of web robots are aware of the existence and purpose of robots.txt files, as these files are a standard and widely recognized means of communicating site crawling preferences. As such, the presence of a disallow directive within the file constitutes strong evidence that any access contrary to these directives is unauthorized and occurs without the webmaster's consent.

The second element involves proving that the defendant's actions interfered with the plaintiff's possessory interest in their digital property—specifically, the website and server resources. When a web robot ignores a robots.txt file and accesses parts of the website that are explicitly disallowed, this constitutes interference. The robots.txt file clearly communicates which areas of the site are off-limits to robots, and the robot's actions in accessing these areas amount to unauthorized use of the website's resources. While the interference occurs in the digital realm, the servers hosting the website are physical assets, and the unauthorized use of these resources can be equated to the physical interference or intermeddling commonly associated with traditional trespass to chattels cases.

The third element to establish is that the defendant intentionally performed the actions constituting the trespass. The operation of a web robot is not an accidental act. Intent, in this context, does not require an intent to cause harm. If a robot accesses restricted parts of a website in violation of the directives laid out in the robots.txt file, it is an intentional act of accessing those areas. The intent refers to the deliberate act of sending the robot to the site and programming it in such a way that it disregards the robots.txt directives. Even if the robot's operator did not specifically aim to defy the robots.txt file, operating a robot that accesses disallowed areas may be considered reckless or indicative of willful

[97] *Grogan v Robin Meredith Plant Hire* [1996] CLC 1127 (CA).
[98] *Donaldson v Becket* (1774) 1 ER 837 (HL).
[99] To establish copyright infringement, the plaintiff must demonstrate both copying of the work (or a substantial portion thereof) and access to the copyright-protected work. Whether a substantial part of the work has been copied is a question of fact. See *Kantel v Grant* [1933] Ex CR 84, 84); *Hawkes & Son Ltd v Paramount Film Service Ltd* [1934] Ch 593.
[100] *eBay, Inc v Bidder's Edge, Inc* (2000) 100 F Supp 2d 1058 (ND Cal).
[101] George E Woodbine, 'The Origins of the Action of Trespass' (1925) 34 The Yale Law Journal 343.
[102] Interestingly, prior to this claim in trespass to chattels, eBay originally filed a copyright infringement claim but it was dismissed before this claim, showcasing the difficulty in establishing a successful copyright claim under this type circumstances. In case a copyright claim exist, trespass to chattels will be preempted by federal Copyright Act in the US.
[103] Laura Quilter, 'The Continuing Expansion of Cyberspace Trespass to Chattels' (2002) 17 Berkeley Tech. LJ 421.
[104] ibid.

ignorance—both of which satisfy the intent requirement under tort law.

The fourth and final element, harm, can often be the most challenging to prove. In this context, harm does not need to involve extensive physical damage. It can include increased server load, which may slow down the website for legitimate users, additional costs incurred for bandwidth and server maintenance due to the increased load, potential security risks or breaches if the robot accesses sensitive areas, or the loss of control over website data and its intended use. However, as Schellekens points out, the level of harm required to warrant a finding of trespass to chattels is highly contentious.[105] In *eBay, Inc. v Bidder's Edge*, the court grappled with whether BE's actions alone constituted sufficient impairment to the value or condition of eBay's system to justify a finding of trespass to chattels. To address this, the court adopted a "slippery slope"[106] argument. It reasoned that while BE used only a small portion of eBay's server capacity, its actions still deprived eBay of the ability to use that portion of its property for its own purposes. The court further argued that failing to hold BE liable could encourage similar behavior by others, potentially overloading eBay's server to the point of causing substantial harm.

If there is no substantial harm to the plaintiff's property, it becomes challenging to establish trespass to chattels. Schellekens argues that this is because trespass to chattels fundamentally falls short—it cannot address harm done to anything other than the plaintiff's physical property, and the evaluation of harm is often arbitrary. He further contends that a more effective framework, consisting of self-regulatory mechanisms and governmental regulation, should be introduced to manage and regulate the behavior of web robots.[107]

While his argument is not without merit, we propose a simpler and more direct legal solution within the existing framework of tort law that addresses the shortcomings he highlights. Specifically, our solution is for courts to clarify that ignoring a robots.txt file and deploying robots to scrape a plaintiff's website without authorization would expose the defendant to liability under the tort of negligence. This would apply in cases where the harm is not directly inflicted on the plaintiff's property but instead on the plaintiff themselves, such as reputational harm or consequential economic loss.

To succeed in a negligence claim, the plaintiff must demonstrate that the defendant owed them a duty of care, breached the applicable standard of care, and caused harm that was not too remote. The key issue is for courts to establish that the deployer of robots owes the webmaster a duty of care. Although there is no direct precedent, this duty can be established by considering the principles set out in the *Caparo Industries PLC v Dickman* case.[108] The courts could find that the harm caused by the robots' actions is reasonably foreseeable and that the webmaster relies on the deployer's adherence to the directives outlined in the robots.txt file. The relationship between the two parties is sufficiently close because the deployer's actions directly impact the webmaster's ability to control access to their website.

Furthermore, it would be fair, just, and reasonable to impose such a duty, as demonstrated by the findings in Section 2.3. We show that webmasters are increasingly updating their robots.txt files to block generative AI bots, reflecting growing concern over unauthorized data extraction. This shift, while understandable, risks fragmenting the internet. It is turning an open, shared space into a patchwork of restricted silos. Smaller websites, lacking the infrastructure to handle high-volume scraping, are disproportionately affected, facing increased costs and service disruptions. Imposing duty of care would help restore balance. It would deter harmful practices, encourage fairer engagement between AI developers and content providers, and ensure that the burdens of AI development are not unfairly placed on those least equipped to bear them.

Once the duty of care is established, the plaintiff must demonstrate that the deployer breached that duty. The expected standard of care is assessed based on whether a reasonable person in the defendant's position would have acted similarly. Since it is a widely recognized community norm to respect and follow the directives outlined in a robots.txt file, a reasonable person would not scrape a website without authorization. Therefore, it is evident that the deployer fell short of the expected standard of care.

The next step is to establish both factual and legal causation between the harm suffered by the webmaster and the deployer's wrongdoing. For factual causation, the "but-for" test applies: would the harm have occurred but for the defendant's breach of duty? The answer is clearly no If the deployer had not deployed the robots to scrape the website, the harm to the webmaster would not have occurred. For example, in an ongoing lawsuit[109] between the New York Times (NYT) and OpenAI, the NYT alleged reputational harm, among others, against OpenAI. When a user asked Bing Chat (which incorporates AI capabilities developed by OpenAI) to summarize an NYT article titled "Dietary Approaches for Heart Health" and provide a list of "15 foods best for heart health," Bing Chat generated a list of foods that included consumption of red wine. However, the original NYT article mentioned none of the foods listed in Bing Chat's response, except for three, and explicitly stated that red wine was not beneficial for heart health. Such inaccuracies, attributed to the AI's training process, raised concerns about the misrepresentation of NYT's content and potential harm to its credibility. In this example, if OpenAI had not scraped articles from NYT and used them to train ChatGPT, the model would not output fabricated content that could damage NYT's reputation.

For legal causation, the question is whether the defendant's breach was the legal cause of the plaintiff's harm. As long as no subsequent event breaks the chain of causation and the harm is not too remote,[110] the causal link is established. In the case of NYT, there is no intervening third party whose actions are wholly unreasonable and unforeseeable, which would break the chain of causation. Moreover, the harm caused—ChatGPT generating misleading information and associating it with NYT, thereby damaging its reputation—was reasonably foreseeable as a result of the deployer's carelessness.

In cases of negligence, courts generally allow recovery for consequential economic losses arising from harm to the plaintiff's person or property. For instance, in the case of NYT, the company's reputation and brand trust may be directly harmed, resulting in consequential economic losses. These losses could include reduced sales, a declining customer base, and missed business opportunities caused by the damage to its reputation. By recognizing and addressing such harms, courts can provide a pathway to recovery for plaintiffs impacted by breaches of duty in the digital age. It is also worth noting that harms to physical property, such as those seen in *eBay v Bidder's Edge*, where computer systems were damaged or burdened, can similarly be recovered under the umbrella of negligence. In this sense, trespass to chattels, as discussed in this context, may be viewed as a specific application of the broader tort of negligence. When the only harm suffered is to physical property, a negligence claim often aligns with trespass to chattels due to the presence of direct precedents. This overlap highlights how negligence principles can provide a comprehensive framework for addressing both physical property harm and intangible harms caused by breaches of duty.

For pure economic loss, which refers to financial loss that does not arise from damage to the plaintiff's person or property, courts are

[105] Schellekens (n 12) 671.

[106] ibid 670.

[107] ibid 674.

[108] *Caparo Industries PLC v Dickman* (1990) 2 AC 605.

[109] Michael M. Grynbaum and Ryan Mac, 'The Times Sues OpenAI and Microsoft over A.I. Use of Copyrighted Work' <https://www.nytimes.com/2023/12/27/business/media/new-york-times-open-ai-microsoft-lawsuit.html> accessed 10 April 2024.

[110] *Page v Smith* [1996] AC 155 (HL).

generally reluctant to recognize claims.[111] An exception was established in *Hedley Byrne v Heller*,[112] which addressed negligent misstatements leading to pure economic loss. However, this exception most likely does not apply, as there is rarely misstatements nor misrepresentation involved in the interaction between the webmaster and the deployer of the web robots. To better understand how pure economic loss may arise in the context of web scraping, let us consider the following hypothetical example: the plaintiff regularly uploads mathematical proofs of complex problems to her blog. The defendant violates plaintiff's robots.txt directives and terms of use by scraping the blog, compiling the proofs into a book, and selling it for profit. In this scenario, no contract exists between plaintiff and defendant, as there is no meeting of minds. Additionally, copyright law offers no remedy because mathematical proofs are not copyrightable. Furthermore, there is no direct harm to plaintiff's blog server, meaning trespass to chattels does not apply, and there is no personal injury to plaintiff. The harm, if any, is purely economic—the loss of opportunity for plaintiff to earn money by selling the book herself. Whether such behavior should give rise to legal liability is a matter of policy. Even if an exception for pure economic loss were considered, the harm in this case would likely be deemed too remote to establish causation. This example underscores the challenges of addressing pure economic loss in negligence claims and highlights the need for thoughtful consideration of policy implications in such cases.

To conclude, it is evident that both trespass to chattels and negligence offer vital frameworks for addressing the unauthorized use of web robots, each serving distinct yet complementary roles when other causes of actions are not available. Trespass to chattels is effective for cases involving direct interference with physical property, such as server overload, while negligence extends its reach to encompass harm to individuals, reputations, and consequential economic losses.

From a policy perspective, adapting tort principles to regulate web scraping is not just a matter of legal redress but a necessary step toward ensuring equity and accountability in the digital age. We have shown that rapid evolution of generative AI and large-scale data scraping practices has disrupted the cooperative balance of the internet, creating significant disparities between powerful AI developers and smaller content providers. By recognizing a duty of care and imposing liability on those who disregard robots.txt directives, courts can promote responsible behavior, discourage exploitative practices, and reduce the negative social impacts of unchecked scraping. This approach not only safeguards the rights of webmasters but also fosters a more sustainable and inclusive digital ecosystem, ensuring that innovation does not come at the expense of fairness and collaborative values.

## 5. Conclusion

The analysis presented in this paper highlights the compelling role of robots.txt at the intersection of technology, law, and norms. Contrary to the common belief held by many tech researchers, that violating robots.txt leads only to ethical concerns, this paper demonstrates that, under certain circumstances, violations of restrictions outlined in robots.txt can lead to legal liabilities. Through a thorough exploration of legal theories in the domains of contract law and tort law, it provides insights into the legal recourse available when robots.txt violations occur, depending on specific contexts and jurisdictions.

Although this paper limits its analytical scope to common law systems, we believe its findings offer generalizable insights. As a globally adopted technical standard, robots.txt presents legal challenges that transcend jurisdictional boundaries. Much like how the EU's General Data Protection Regulation (GDPR),[113] despite being jurisdiction-specific, has shaped privacy laws in diverse legal systems such as China's Personal Information Protection Law (PIPL)[114] and Kenya's data protection regime,[115] we hope this analysis can contribute to broader discussions on digital governance and automated access regulation in the age of AI.

From a broader perspective, our insights can also be extended to several related areas. First, the rise of generative AI and their reliance on extensive data necessitate a reevaluation of the balance between innovation and digital equity. The legal frameworks proposed here serve as a starting point for mitigating the negative impacts of unregulated scraping. Second, these insights have direct relevance to the development of international legal standards and policies, particularly as they pertain to AI ethics and governance. The trajectory of robots.txt regulation is likely to evolve in tandem with advancements in AI and data-intensive technologies. Courts, policymakers, and industry leaders must collectively address the vulnerabilities identified in this paper by establishing clearer legal guidelines, promoting ethical practices, and ensuring equitable access to digital resources.

Ultimately, the challenges surrounding robots.txt reflect broader tensions between technological progress, legal structures, and societal values. Addressing these tensions will require an interdisciplinary approach, combining technological solutions with robust legal and policy frameworks. By doing so, we can preserve the foundational principles of the internet—openness, collaboration, and mutual benefit—while adapting to the demands of an increasingly data-driven world. In a broader societal sense, clearer approaches can help foster fair competition, protect smaller data holders, and ensure that the benefits of data-driven innovation are more widely shared. They can also strengthen public trust by showing that automated data access is governed by not only technological constraints, but by clear-cut legal and ethical standards.

[111] *Spartan Steel & Alloys Ltd v Martin & Co Ltd* [1973] QB 27 (CA).

[112] *Hedley Byrne & Co Ltd v Heller & Partners Ltd* [1964] AC 465.

[113] Regulation (EU) 2016/679 of the European Parliament and of the Council of 27 April 2016 on the protection of natural persons with regard to the processing of personal data and on the free movement of such data (General Data Protection Regulation) [2016] OJ L119/1.

[114] Daniel Albrecht, 'Chinese First Personal Information Protection Law in Contrast to the European GDPR' (2022) 23 Computer Law Review International 1.

[115] Eugene Wanekeya, 'Effectiveness of Domestic Data Protection Laws in African Countries: A Case Study of the Data Protection Law in Kenya' (University of Nairobi 2023).

**Declaration of competing interest**

The authors declare that they have no known competing financial interests or personal relationships that could have appeared to influence the work reported in this paper.

**Data availability**

Data will be made available on request.